\documentclass[10pt,journal,compsoc]{IEEEtran}
\usepackage[T1]{fontenc}
\usepackage{amsmath,amssymb,amsthm}
\usepackage{graphicx}
\usepackage{booktabs,array}
\usepackage{microtype}
\usepackage{xurl}
\usepackage[hidelinks]{hyperref}
\hypersetup{
  pdftitle={Signed but Unbound: The Authority Gap Between Agent Commerce Mandates and Crypto Wallet Signatures},
  pdfauthor={Shengwei You, Aditya Joshi, Andrey Kuehlkamp, and Jarek Nabrzyski},
  pdfkeywords={Agentic commerce, authorization, cryptocurrency wallets, EIP-3009, smart contracts, stablecoins, x402}
}
\usepackage{tikz}
\usetikzlibrary{arrows.meta,positioning,fit,backgrounds}
\graphicspath{{figures/}}
\newcommand{\nEvents}{26,653}
\newcommand{\nNAuth}{23,156}
\newcommand{\nDecodeRate}{86.9\%}
\newcommand{\nWinSampled}{60}
\newcommand{\nWinMinSec}{20}
\newcommand{\nWinMaxSec}{198}
\newcommand{\nDateFrom}{4 May 2026}
\newcommand{\nDateTo}{16 November 2026}

\newcommand{\nPathTransfer}{86.8\%}

\newcommand{\nPathBatcher}{11.9\%}
\newcommand{\nPathCPP}{0.5\%}
\newcommand{\nPathOther}{0.7\%}
\newcommand{\nNBatcher}{3,178}
\newcommand{\nNCPP}{136}
\newcommand{\nCommitShare}{12.4\%}

\newcommand{\nBearerShare}{98.7\%}
\newcommand{\nDirectShare}{93.5\%}
\newcommand{\nDirectThird}{99.5\%}

\newcommand{\nXAttributed}{22,655}
\newcommand{\nXShare}{97.8\%}
\newcommand{\nXFacs}{14}
\newcommand{\nXCoinbase}{92.5\%}
\newcommand{\nXReceive}{0.0\%}
\newcommand{\nXCommit}{5732bb8}

\newcommand{\nResMedian}{296}

\newcommand{\nResGtHour}{0.8\%}

\newcommand{\nResGtMonth}{0.6\%}
\newcommand{\nValidAfterZero}{21.2\%}
\newcommand{\nValueMedian}{0.0055}

\newcommand{\nValueLtDollar}{99.0\%}
\newcommand{\nSliceRecords}{2,674}
\newcommand{\nConcurrent}{40.1\%}
\newcommand{\nSliceSeconds}{20}
\newcommand{\nSliceAll}{3,436}
\newcommand{\nSliceBearer}{3,378}
\newcommand{\nSliceBearerShare}{98.3\%}
\newcommand{\nOtherDecoded}{2,205}
\newcommand{\nOtherBearer}{2,178}
\newcommand{\nOtherBearerShare}{98.8\%}
\newcommand{\nConfigs}{12}
\newcommand{\nCells}{108}
\newcommand{\nCellsExec}{106}
\newcommand{\nNTrials}{31,800}

\newcommand{\nNMandates}{300}
\newcommand{\nNHonest}{3,600}

\newcommand{\nSweepTotal}{12,000}
\newcommand{\nSweepMutated}{10,800}
\newcommand{\nSweepMutAccepted}{0}
\newcommand{\nSweepUnmodified}{1,200}
\newcommand{\nSweepAccepted}{233}
\newcommand{\nSweepMismatch}{0}
\newcommand{\nWalletCast}{50/50}
\newcommand{\nWalletEA}{50/50}

\newcommand{\nGasReceive}{78,588}
\newcommand{\nGasLock}{98,084}
\newcommand{\nGasCPP}{151,358}
\newcommand{\nGasHardened}{91,408}

\newcommand{\nGasOverheadPct}{25\%}
\newcommand{\nBaseFeeGwei}{0.005}
\newcommand{\nEthUsd}{1,819.45}
\newcommand{\nCostTransfer}{0.00072}
\newcommand{\nCostLock}{0.00090}
\newcommand{\nCostCPP}{0.00138}

\newcommand{\nCostDelta}{0.00018}
\newcommand{\nCostLockPctMedian}{16\%}
\newcommand{\nCostTransferPctMedian}{13\%}



\newtheorem{definition}{Definition}
\newtheorem{proposition}{Proposition}
\newtheorem{theorem}{Theorem}
\newcommand{\para}[1]{\smallskip\noindent\textbf{#1.}\ }
\newcommand{\Auth}{\mathsf{Auth}}
\newcommand{\Ex}{\mathsf{Ex}}
\newcommand{\Bind}{\mathsf{Bind}}
\newcommand{\Acc}{\mathsf{Accept}}
\newcommand{\N}{\mathsf{N}}

\newcommand{\ML}{\textsc{MandateLock}}
\newcommand{\cls}[1]{\textup{\textsc{#1}}}
\newcommand{\RWA}{\texttt{receiveWith\allowbreak Authorization}}
\begin{document}
\title{Signed but Unbound: The Authority Gap Between Agent Commerce Mandates and Crypto Wallet Signatures}
\author{Shengwei~You, Aditya~Joshi, Andrey~Kuehlkamp, and Jarek~Nabrzyski%
\IEEEcompsocitemizethanks{\IEEEcompsocthanksitem The authors are with the Department of Computer Science and Engineering, University of Notre Dame, IN 46556, USA.}}
\markboth{Preprint}{Signed but Unbound: The Authority Gap Between Agent Commerce Mandates and Wallet Signatures}

\IEEEtitleabstractindextext{%
\begin{abstract}
When an artificial intelligence agent pays with a stablecoin, the commerce protocol can specify the merchant, order, amount, and facilitator. The token contract enforces the fields the wallet signs. We call the difference the \emph{authority gap}, define nine constraint classes, and derive them for \nConfigs{} authorization configurations, including Coinbase's deployed Commerce Payments Protocol. All \nNTrials{} trials on local copies of deployed contract code match the derivation. A controlled comparison shows that committing the terms in the signed nonce binds the order. When the facilitator submits a payment to a merchant account without code, binding the facilitator requires a separate recipient contract. Of \nEvents{} sampled USD Coin authorizations redeemed on Base, \nBearerShare{} followed a path that any holder of the signature could redeem. We prove sufficient conditions for a recipient contract to enforce the signed mandate under a trusted signer and show which recipient a wallet displays. \ML{}, a minimal implementation for immediate payments, rejects all \nSweepMutated{} modified mandates on deployed token code and accepts signatures from two independent signers. It uses \nGasOverheadPct{} more gas than a transfer that only the recipient can submit.
\end{abstract}
}
\maketitle
\IEEEdisplaynontitleabstractindextext
\IEEEpeerreviewmaketitle

\section{Introduction}
\IEEEPARstart{A}{gent} commerce protocols move payment authority from a person to software. Google's Agent Payments Protocol (AP2) has the user sign \emph{mandates} that state what an agent may buy, from whom and until when~\cite{ap2}. Coinbase's x402 lets an agent pay for a web resource by attaching a signed stablecoin authorization that a third-party \emph{facilitator} submits to the chain~\cite{x402}. In each design, the user's intent is captured at the protocol layer and executed by a payment instrument.

On card rails the instrument has an enforcer: the issuer that minted a scoped token checks its merchant, amount and expiry before money moves~\cite{acp}. On a public blockchain the only enforcer is the token contract, and it knows nothing about mandates. It checks the fields of the structure the wallet signed. Any constraint of the mandate that the signed structure does not encode is enforced, if at all, by whoever holds the signature: the agent runtime, the merchant server, an HTTP intermediary or the facilitator. Recent work shows that these parties can be manipulated or compromised~\cite{li2026fiveattacks,ling2026freeriding,wang2026http402,louck2026whisper,aviv2026beyondmandate}.

This paper studies the boundary between the commerce protocol and the wallet. We ask: which mandate constraints does each wallet authorization mechanism enforce on chain; how is the boundary configured in deployed agent payments; and what is required to make the signature enforce exactly the mandate, without changing deployed tokens?

\para{Approach} We model a mandate as a set of allowed on-chain histories and a signed authorization as the set of histories that any holder of it can cause. Their difference is the \emph{authority gap}. We compare nine classes of violations involving the payee, per-transfer and cumulative spending limits, expiry, redeemer, order, chain, residual authority and revocation (Section~\ref{sec:model}). For each configuration we derive which classes its redemption path binds. We then (i) execute every class against every configuration on runtime bytecode fetched from Base, including Circle's USDC, Uniswap's Permit2 and Coinbase's Commerce Payments Protocol (CPP)~\cite{cpp} (Section~\ref{sec:matrix}); (ii) measure how deployed payments use these mechanisms (Section~\ref{sec:wild}); and (iii) state and prove conditions under which a recipient contract makes the signature mandate-exact (Sections~\ref{sec:design}--\ref{sec:locketal}).

\para{Findings} All \nNTrials{} executed trials agree with the derivation, with outcomes invariant across \nNMandates{} randomized mandates per cell. Two ablation rows isolate the mechanisms. Committing the mandate terms in the signer-chosen EIP-3009 nonce closes order substitution without any contract. Binding the redeemer requires the payee-bound \texttt{receiveWithAuthorization}, whose caller must equal the signed recipient. For a merchant account without code, settlement initiated by a separate facilitator therefore requires a recipient contract. A wallet that renders the signed fields shows that contract as the payee (Proposition~\ref{prop:visibility}). In the Base USDC sample, we identified bearer redemption for \nBearerShare{} of the \nEvents{} authorizations. We identified on-chain nonce derivation in \nCommitShare{} of sampled events, mostly through a batching contract that still uses the bearer variant. CPP, which deploys commitment and payee binding together, accounts for \nPathCPP.

\para{Contributions}
\begin{enumerate}
\item \emph{A model of the agent-commerce/wallet trust boundary} that defines the authority gap and compares nine classes of violations across \nConfigs{} configurations, including deployed and ablated ones (Section~\ref{sec:model}).
\item \emph{Validation on deployed bytecode and a component ablation}, \nNTrials{} trials in \nCellsExec{} executed cells, which separate the effect of nonce commitment, caller binding and recipient contracts, and a visibility trade-off that follows from EIP-3009 semantics (Section~\ref{sec:matrix}).
\item \emph{A measurement of \nEvents{} redeemed authorizations} on Base, with \nXShare{} of decoded ones attributed to registered x402 facilitators, that quantifies bearer redemption, observable on-chain nonce derivation, the withholding horizon and same-terms concurrency (Section~\ref{sec:wild}).
\item \emph{Verifier conditions and a game-based soundness proof} for commitment-bound redemption, a pattern CPP deploys without a published analysis, with \ML{} as a minimal one-shot instance evaluated for exactness, spoofing resistance, signer compatibility and cost (Sections~\ref{sec:design} and~\ref{sec:locketal}).
\end{enumerate}
The reviewer supplement contains the measurement inputs, recorded experiment results, and an offline check of authorized settlement.

\section{Background}
\label{sec:background}
\para{Agent commerce protocols} In \emph{AP2}~\cite{ap2}, an \emph{intent mandate} records what the user asked the agent to do, a \emph{cart mandate} records the cart the merchant offered and the user approved, and a \emph{payment mandate} conveys the authorization to the payment side; a \emph{credentials provider} holds the user's payment methods. AP2 includes an extension for settling through x402. In \emph{x402}~\cite{x402}, a server answers a request with HTTP status 402 and \texttt{PaymentRequirements}: scheme, network, asset, amount, recipient (\texttt{payTo}), resource and timeout. The client returns a signed payload, and a facilitator verifies it and submits it on chain. In the dominant \emph{exact} scheme on Ethereum-compatible chains, the payload is an EIP-3009 \texttt{transferWithAuthorization}. Base is Coinbase's Ethereum layer-2 network (chain identifier 8453) and the main venue of x402 settlement~\cite{wang2026http402}. The OpenAI--Stripe Agentic Commerce Protocol (ACP) delegates a scoped card credential instead~\cite{acp}.

\para{Wallet authorization primitives} An externally owned account (EOA), an account controlled by one private key, can authorize a token movement in several ways. An ERC-20 \texttt{approve} transaction grants a spender an allowance, which the spender draws with \texttt{transferFrom} to any recipient~\cite{erc20}. An ERC-2612 \texttt{permit} is an off-chain signature, in the EIP-712 typed-data format~\cite{eip712}, that sets the same allowance when anyone submits it before a deadline~\cite{eip2612}. EIP-3009~\cite{eip3009} defines a signature over \emph{(from, to, value, validAfter, validBefore, nonce)}. Its \texttt{transferWithAuthorization} may be submitted by anyone; its \texttt{receiveWithAuthorization} uses a different type hash and requires the caller to equal \emph{to}. The 32-byte \emph{nonce} is chosen by the signer, and the token only checks that it is unused. USDC implements EIP-3009 in Circle's FiatToken contract~\cite{fiattoken}. Uniswap's Permit2 \texttt{permitTransferFrom} is a signature over token, amount, spender, nonce and deadline, redeemable only by the spender, which chooses the recipient. Its \texttt{permitWitnessTransferFrom} also hashes a spender-defined \emph{witness}~\cite{permit2}. Permit2 requires the owner to have approved the Permit2 contract, usually for an unlimited amount. Signature primitives fix the chain through the EIP-712 domain separator; transactions fix it through EIP-155~\cite{eip155}.

\para{Commerce Payments Protocol} Coinbase's CPP~\cite{cpp} is an escrow for authorize-and-capture payments. Its \texttt{ERC3009PaymentCollector} calls \texttt{receiveWithAuthorization} with itself as recipient and with the nonce set to a hash of the payment terms (\texttt{PaymentInfo}: operator, receiver, amount, expiries, fee bounds and salt), and its escrow lets only the named operator act. CPP is, to our knowledge, the first deployment of commitment-bound EIP-3009 redemption. We evaluate its deployed v1.1 contracts as a configuration.

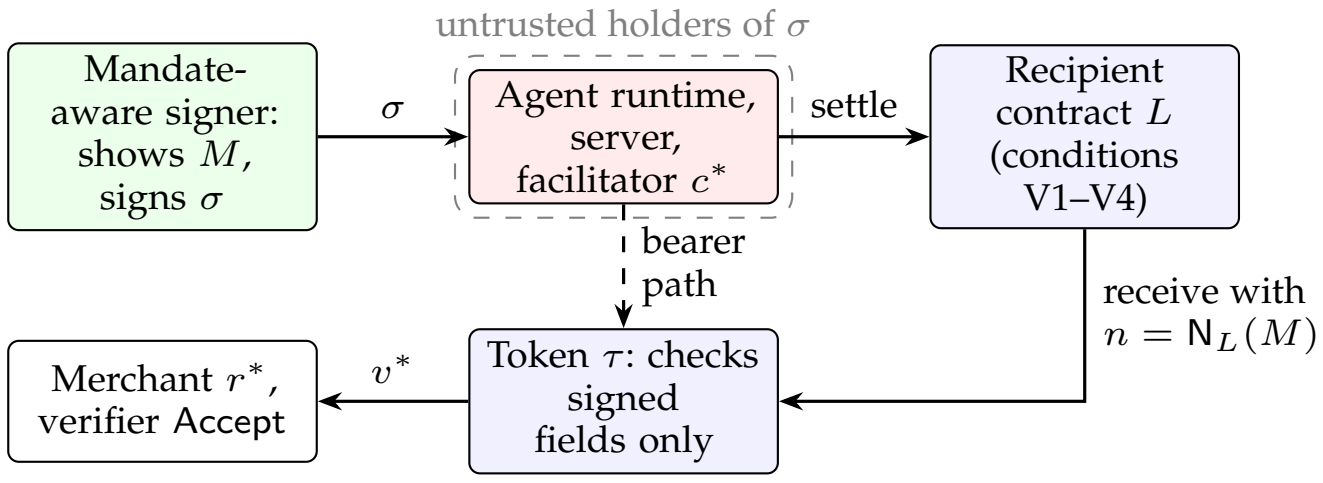
\begin{figure}[t]
\centering
\begin{tikzpicture}[font=\scriptsize,
  box/.style={draw,rounded corners=2pt,align=center,minimum height=8mm,text width=19mm,inner sep=2pt},
  tr/.style={box,fill=green!8},un/.style={box,fill=red!8},ch/.style={box,fill=blue!6},
  ar/.style={-{Stealth[length=1.6mm]},semithick}]
\node[tr] (sig) at (0,0) {Mandate-aware signer:\\shows $M$, signs $\sigma$};
\node[un] (hold) at (3.05,0) {Agent runtime, server,\\facilitator $c^\ast$};
\node[ch] (L) at (6.1,0) {Recipient contract $L$\\(conditions V1--V4)};
\node[ch] (tok) at (3.05,-1.75) {Token $\tau$: checks\\signed fields only};
\node[box] (mer) at (0,-1.75) {Merchant $r^\ast$,\\verifier $\Acc$};
\draw[ar] (sig) -- node[above]{$\sigma$} (hold);
\draw[ar] (hold) -- node[above]{settle} (L);
\draw[ar] (L.south) |- node[pos=0.25,right,align=left]{receive with\\$n=\N_L(M)$} (tok.east);
\draw[ar,dashed] (hold.south) -- node[right,align=left]{bearer\\path} (tok.north);
\draw[ar] (tok.west) -- node[above]{$v^\ast$} (mer.east);
\begin{scope}[on background layer]
\node[draw=gray,dashed,rounded corners,fit=(hold),inner sep=2.5pt,label={[gray,font=\scriptsize]above:untrusted holders of $\sigma$}] {};
\end{scope}
\end{tikzpicture}
\caption{The trust boundary. The commerce protocol states the mandate $M$; the token enforces only the signed fields. In the bearer path (dashed) any holder of $\sigma$ calls the token directly. A recipient contract $L$ checks the redeemer, window and committed terms and forwards $v^\ast$ to the merchant.}
\label{fig:overview}
\end{figure}

\section{Threat Model and Scope}
\label{sec:threat}
\para{Parties} Fig.~\ref{fig:overview} shows the parties. A \emph{payer} $p$ holds a code-less EOA key. A \emph{merchant} $r^\ast$ quotes a price for an order. A \emph{facilitator} $c^\ast$ submits settlement and pays its gas. The \emph{agent runtime}, the \emph{merchant server}, HTTP intermediaries and any third party may observe the signed authorization in transit.
\para{Trusted base} We trust the payer's key, the token and Permit2 as deployed, and the component that turns a mandate into the signed message. In AP2 this component is the credentials provider; otherwise it is a mandate-aware wallet or wallet extension. It shows the user the mandate and computes any commitment the configuration requires. Section~\ref{sec:visibility} explains why this trust is necessary for commitment-bound configurations.
\para{Adversary} Every other party may be adversarial. The adversary sees all signed authorizations, controls arbitrary accounts and contracts, and orders transactions. For the \cls{payee}, \cls{overdraw}, \cls{total}, \cls{late}, \cls{order}, \cls{chain}, \cls{residual} and \cls{revoke} classes, it also controls the key of the party the configuration designates as redeemer, which models a compromised facilitator. For \cls{redeemer} it is any other account. It cannot forge signatures or find Keccak-256 collisions.
\para{Scope} We study single payments of an exact amount, as in x402's exact scheme and AP2 cart mandates. Mandates that allow variable amounts up to a cap, recurring payments and refunds are outside the model. Manipulating the agent \emph{before} the mandate is signed is a separate problem with its own defenses~\cite{louck2026whisper,debi2026whispers,aviv2026beyondmandate,camel}. We ask whether the chain enforces whatever mandate was signed, and nothing more. Whether the merchant then delivers the resource is also out of scope~\cite{a402,asp2026}.

\section{The Authority Gap}
\label{sec:model}
\subsection{Mandates, Histories and Attribution}
\begin{definition}[Mandate]
A mandate is $M=(\chi,\tau,p,r^\ast,v^\ast,t_0,t_1,c^\ast,o^\ast,h)$: chain $\chi$, token $\tau$, payer $p$, merchant $r^\ast$, exact amount $v^\ast$, window $[t_0,t_1]$, designated redeemer $c^\ast$ (or $\ast$ for any), order identifier $o^\ast$ and cart or request digest $h$.
\end{definition}
An x402 \texttt{PaymentRequirements} object or an AP2 cart mandate determines $M$: \texttt{payTo} or the merchant gives $r^\ast$, the price gives $v^\ast$, the timeout or expiry gives $t_1$, the facilitator gives $c^\ast$, and the resource or cart gives $o^\ast$ and $h$.

A \emph{history} is a sequence of transactions. We attribute token movements to an authorization when a call consumes it, forwards its proceeds, or draws on an allowance it created. Its \emph{net effect} sums these balance changes per account. Thus a receipt and immediate forward count as one payment even when the same transaction batches other payments. The merchant's \emph{verifier} reads the on-chain evidence for the redemption under review:
\begin{multline*}
\Acc_{\mathcal{K}}(o,\text{logs})=\mathsf{paid}(r^\ast,v^\ast)\\
\wedge\big(\kappa_{\mathcal{K}}(o)=\bot\vee\kappa_{\mathcal{K}}(o)\in\text{logs}_{E_{\mathcal{K}}}\big).
\end{multline*}
Here $\mathsf{paid}$ requires a $\tau$ transfer of $v^\ast$ to $r^\ast$ in that redemption; $\kappa_{\mathcal{K}}(o)$ is the order commitment of configuration $\mathcal{K}$, which the merchant recomputes from its own quote; and $\text{logs}_{E_{\mathcal{K}}}$ are the topics emitted by the configuration's contract $E_{\mathcal{K}}$ for that redemption. Configurations without an order commitment have $\kappa=\bot$: the merchant can match payee and amount, and must rely on off-chain bookkeeping of nonces to tell orders apart.

\begin{definition}[Authorized and redeemable histories]
Let $\sigma$ be the authorization the payer's signer produced for $M$ under configuration $\mathcal{K}$. $\Auth(M)$ is the set of histories in which $\sigma$ either causes no movement of funds or settles in one transaction whose attributable net effect is one transfer of exactly $v^\ast$ from $p$ to $r^\ast$ on $(\chi,\tau)$. That transfer is caused by $c^\ast$ (any account if $c^\ast=\ast$), occurs at a time in $[t_0,t_1]$ before any revocation by $p$, and satisfies $\Acc(o,\cdot)$ only for $o=o^\ast$. $\Auth_{\mathcal{K}}(\sigma)$ is the set of histories the adversary of Section~\ref{sec:threat} can cause the chain to accept.
\end{definition}
\begin{definition}[Authority gap]
$\Ex(M,\sigma)=\Auth_{\mathcal{K}}(\sigma)\setminus\Auth(M)$; $\sigma$ is \emph{mandate-exact} if $\Ex(M,\sigma)=\emptyset$.
\end{definition}

\subsection{Nine Constraint Classes}
Tables~\ref{tab:binding} and~\ref{tab:matrix} compare nine possibly overlapping classes of violations.
\begin{itemize}
\item \cls{payee}: a transfer reaches a recipient other than $r^\ast$.
\item \cls{overdraw}: one transfer moves more than $v^\ast$.
\item \cls{total}: the transfers together move more than $v^\ast$.
\item \cls{late}: a transfer occurs after $t_1$.
\item \cls{redeemer}: a transfer is caused by an account other than $c^\ast$.
\item \cls{order}: the verifier accepts the history as payment for an order $o\neq o^\ast$ of equal price.
\item \cls{chain}: a transfer occurs on a chain other than $\chi$.
\item \cls{residual}: after the mandated transfer, $\Auth_{\mathcal{K}}(\sigma)$ still contains histories with further transfers from $p$.
\item \cls{revoke}: a transfer occurs after $p$ revoked by the configuration's own mechanism.
\end{itemize}
A mandate for ten tokens permits one payment of ten or no payment. An ERC-20 allowance of ten permits two separate transactions of five each~\cite{erc20}. Both draws respect the per-transfer and cumulative limits, but the history lies outside $\Auth(M)$. The nine columns alone therefore cannot establish mandate exactness. Likewise, \cls{late} checks expiry only. Exactness also requires settlement at or after $t_0$.

A configuration $\mathcal{K}$ \emph{binds} class $k$ if its token-side and contract-side checks exclude every history of class $k$ from $\Auth_{\mathcal{K}}(\sigma)$. Let $\Bind(\mathcal{K})$ be the set of bound classes.

\begin{proposition}[Derivation]
\label{prop:derivation}
Fill the signed fields from $M$ as listed in Appendix~\ref{app:derivation}: for example, EIP-3009 validity ends at $t_1$, and Permit2's deadline is $t_1$. Then, under the semantics of the respective specifications and contracts, $\Bind(\mathcal{K})$ is as given in Table~\ref{tab:binding} for each configuration. For every class outside $\Bind(\mathcal{K})$, Appendix~\ref{app:derivation} gives a history in $\Ex(M,\sigma)$.
\end{proposition}
\begin{proof}[Proof sketch]
For each row and class, either a check on the redemption path compares a signed or committed field with the constraint, which excludes the class, or no check does, in which case the appendix gives a history that the contracts accept. For example, \texttt{transferWithAuthorization} checks $(p,r^\ast,v^\ast)$ and the window, and marks the nonce used. It never reads the caller, so a non-designated submitter gives a \cls{redeemer} history. With a random nonce, an equal-priced second order gives an \cls{order} history. Section~\ref{sec:matrix} executes every witness.
\end{proof}

\begin{table*}[!t]
\caption{Classes bound by each configuration (\checkmark{} bound, $\times$ open). Both ERC-20 approval rows and ERC-2612 permit partial draws against an allowance. Direct Permit2 allows one draw up to the signed limit. Each remaining row requires the full signed amount in one redemption. \emph{Redeemer}: who may submit settlement. \emph{Standing}: the payer keeps a nonzero allowance to a third party after settlement. \emph{Payee in request}: the merchant appears in the typed data or transaction presented for signing. No row changes the token. $C(M)$: nonce set to a hash of the mandate terms.}
\label{tab:binding}
\centering\footnotesize
\setlength{\tabcolsep}{3pt}
\resizebox{\textwidth}{!}{%
\begin{tabular}{@{}lccccccccc|ccc@{}}
\toprule
Configuration & \cls{payee} & \cls{overdraw} & \cls{total} & \cls{late} & \cls{redeemer} & \cls{order} & \cls{chain} & \cls{residual} & \cls{revoke} & Redeemer & Standing & Payee in request \\
\midrule
ERC-20 \texttt{approve}, exact & $\times$ & \checkmark & \checkmark & $\times$ & \checkmark & $\times$ & \checkmark$^{c}$ & \checkmark & \checkmark$^{d}$ & spender & no & no \\
ERC-20 \texttt{approve}, unlimited & $\times$ & $\times$ & $\times$ & $\times$ & \checkmark & $\times$ & \checkmark$^{c}$ & $\times$ & \checkmark$^{d}$ & spender & yes & no \\
ERC-2612 \texttt{permit} & $\times$ & \checkmark & \checkmark & $\times$ & \checkmark & $\times$ & \checkmark & \checkmark & \checkmark$^{d}$ & spender & no & no \\
EIP-3009 transfer (x402 \emph{exact}) & \checkmark & \checkmark & \checkmark & \checkmark & $\times$ & $\times$ & \checkmark & \checkmark & \checkmark$^{d}$ & any & no & yes \\
EIP-3009 receive$^{a}$ & \checkmark & \checkmark & \checkmark & \checkmark & \checkmark & $\times$ & \checkmark & \checkmark & \checkmark$^{d}$ & payee & no & yes \\
Permit2 transfer & $\times$ & \checkmark & \checkmark & \checkmark & \checkmark & $\times$ & \checkmark & \checkmark & \checkmark$^{d}$ & spender & yes & no \\
Permit2 witness proxy (x402-style) & \checkmark & \checkmark & \checkmark & \checkmark & $\times$ & $\times$ & \checkmark & \checkmark & \checkmark$^{d}$ & any & yes & yes \\
Permit2 hardened proxy$^{b}$ & \checkmark & \checkmark & \checkmark & \checkmark & \checkmark & \checkmark & \checkmark & \checkmark & \checkmark$^{d}$ & facilitator & yes & yes \\
\midrule
EIP-3009 transfer + $C(M)$ nonce & \checkmark & \checkmark & \checkmark & \checkmark & $\times$ & \checkmark & \checkmark & \checkmark & \checkmark$^{d}$ & any & no & yes \\
EIP-3009 receive + $C(M)$ nonce$^{a}$ & \checkmark & \checkmark & \checkmark & \checkmark & \checkmark & \checkmark & \checkmark & \checkmark & \checkmark$^{d}$ & payee & no & yes \\
\midrule
CPP \texttt{charge} (deployed) & \checkmark & \checkmark & \checkmark & \checkmark & \checkmark & \checkmark & \checkmark & \checkmark & \checkmark$^{d}$ & facilitator & no & no \\
\ML{} & \checkmark & \checkmark & \checkmark & \checkmark & \checkmark & \checkmark & \checkmark & \checkmark & \checkmark$^{d}$ & facilitator & no & no \\
\bottomrule
\end{tabular}}
\\[2pt]\raggedright\footnotesize $^{a}$In these rows the merchant account has no code and submits its own transaction ($c^\ast=r^\ast$). $^{b}$Our composite of the facilitator-bound settlement of~\cite{li2026fiveattacks} and the request binding of~\cite{ling2026freeriding}, as a Permit2 witness. $^{c}$Through EIP-155 transaction binding, not a token check. $^{d}$Only if the revocation is included before settlement.
\end{table*}

\subsection{Caller Binding and Payee Visibility}
\label{sec:visibility}
Table~\ref{tab:binding} shows that none of the configurations combines an empty gap, zero standing allowance, facilitator-paid settlement and a signing request that names the merchant. For EIP-3009 the merchant account's ability to execute code matters.
\begin{proposition}[Visibility]
\label{prop:visibility}
Suppose a designated facilitator distinct from a merchant with a code-less account initiates the settlement transaction. If redemption of an EIP-3009 authorization delivers the payment to the merchant in that transaction and binds \cls{redeemer}, the signed recipient must be a separate contract.
\end{proposition}
\begin{proof}
The transfer variant accepts any caller. The receive variant requires the signed recipient to call the token. A code-less merchant cannot make that call inside the facilitator's transaction. Naming the facilitator's own code-less account would pay the facilitator. Delivering the proceeds to the merchant in the same redemption therefore requires a contract that calls the token and forwards the payment.
\end{proof}
For this route the EIP-712 fields name the forwarding contract $L$. In \ML{} the nonce carries the order commitment. A merchant that operates its own receiving contract can instead check the facilitator and call the token at the merchant address; its address then remains the signed recipient. Commitment-bound configurations need a signer that shows the mandate and computes its commitment. We put that signer in the trusted base (Section~\ref{sec:threat}). Permit2 witness proxies avoid the visibility loss, because the witness can name the merchant, but they require a standing Permit2 allowance, which is the authority that Permit2 signature phishing abuses.

\section{Validation on Deployed Bytecode}
\label{sec:matrix}
\subsection{Setup}
\para{Contracts} We fetched the runtime bytecode of USDC on Base (FiatToken v2.2 implementation and its linked signature-checking library), Permit2, and CPP v1.1 (escrow, ERC-3009 collector and token-store implementation) at Base block 58,495,000. The supplement records the USDC contract addresses and SHA-256 digests of the included bytecode. The harness runs in Foundry~1.7.1~\cite{foundry} with chain identifier 8453. It places each runtime code at its mainnet address; for USDC, the implementation code replaces the proxy, and the token is set up through its own initializers. The harness checks that the resulting EIP-712 domain separators of USDC and Permit2 equal the values the live contracts return. No transaction is sent to a live network.
\para{Configurations} We use the \nConfigs{} rows of Table~\ref{tab:binding}. The two Permit2 proxies are our implementations; all other rows run deployed or standard code paths. The x402 baseline draws a random nonce, as x402 clients do. The $C(M)$ rows set the nonce to a hash of the mandate terms. Signed windows end at $t_1$, the setting most favorable to each primitive.
\para{Trials} For each configuration we draw \nNMandates{} mandates with pseudorandom keys, amounts between $10^{-6}$ and 1,000 USDC and windows between 30 seconds and one day. From one snapshot per mandate we run an honest settlement and nine attacks, one per class:
\begin{itemize}
\item \cls{payee}: redirect to the attacker;
\item \cls{overdraw}: draw $v^\ast+1$;
\item \cls{total}: attempt another draw after settling the full amount;
\item \cls{late}: redeem 61 seconds after $t_1$, after any innocuous in-window step the primitive allows, such as submitting a permit;
\item \cls{redeemer}: settle from a non-designated account;
\item \cls{order}: settle claiming an equal-priced second order, scored by the verifier $\Acc$;
\item \cls{chain}: redeem under chain identifier 10;
\item \cls{residual}: draw again 30 days after expiry;
\item \cls{revoke}: redeem after the payer's own revocation.
\end{itemize}
An honest settlement counts as exact only if the payer is debited exactly $v^\ast$, the merchant is credited exactly $v^\ast$ and the verifier accepts the logs as payment for $o^\ast$. EIP-155 binds approvals to the chain at the transaction layer, so the two \cls{chain} cells of the approval rows are not executed.

\subsection{Results}
Table~\ref{tab:matrix} reports the outcomes. Of the \nCells{} cells, \nCellsExec{} were executed, for \nNTrials{} trials. Every trial matched Proposition~\ref{prop:derivation}, and within every cell all \nNMandates{} mandates gave the same outcome. All \nNHonest{} honest settlements were exact. Because outcomes do not vary within a cell, the evidence is best read as \nCellsExec{} conformance checks of deployed code against the derived semantics, each replicated over randomized inputs.

\begin{table*}[!t]
\caption{Executed matrix on bytecode fetched from Base: successful attacks out of \nNMandates{} mandates per cell. $\circ$: not executed (EIP-155). Gas: median per-transaction gas of honest settlement in an isolated benchmark (Section~\ref{sec:cost}); approvals and permits need a separate payer transaction and are omitted.}
\label{tab:matrix}
\centering\footnotesize
\setlength{\tabcolsep}{3pt}
\resizebox{\textwidth}{!}{%
\begin{tabular}{@{}lccccccccc|ccr@{}}
\toprule
Configuration & \cls{payee} & \cls{overdraw} & \cls{total} & \cls{late} & \cls{redeemer} & \cls{order} & \cls{chain} & \cls{residual} & \cls{revoke} & Redeemer & Standing & Gas \\
\midrule
ERC-20 \texttt{approve}, exact & \textbf{300} & 0 & 0 & \textbf{300} & 0 & \textbf{300} & $\circ$ & 0 & 0 & designated & no & -- \\
ERC-20 \texttt{approve}, unlimited & \textbf{300} & \textbf{300} & \textbf{300} & \textbf{300} & 0 & \textbf{300} & $\circ$ & \textbf{300} & 0 & designated & yes & -- \\
ERC-2612 \texttt{permit} & \textbf{300} & 0 & 0 & \textbf{300} & 0 & \textbf{300} & 0 & 0 & 0 & designated & no & -- \\
EIP-3009 transfer (x402 \emph{exact}) & 0 & 0 & 0 & 0 & \textbf{300} & \textbf{300} & 0 & 0 & 0 & any & no & 78,553 \\
EIP-3009 receive$^{a}$ & 0 & 0 & 0 & 0 & 0 & \textbf{300} & 0 & 0 & 0 & payee & no & 78,588 \\
Permit2 transfer & \textbf{300} & 0 & 0 & 0 & 0 & \textbf{300} & 0 & 0 & 0 & designated & yes & 81,829 \\
Permit2 witness proxy (x402-style) & 0 & 0 & 0 & 0 & \textbf{300} & \textbf{300} & 0 & 0 & 0 & any & yes & 89,796 \\
Permit2 hardened proxy$^{b}$ & 0 & 0 & 0 & 0 & 0 & 0 & 0 & 0 & 0 & designated & yes & 91,408 \\
\midrule
EIP-3009 transfer + $C(M)$ nonce & 0 & 0 & 0 & 0 & \textbf{300} & 0 & 0 & 0 & 0 & any & no & 78,553 \\
EIP-3009 receive + $C(M)$ nonce$^{a}$ & 0 & 0 & 0 & 0 & 0 & 0 & 0 & 0 & 0 & payee & no & 78,580 \\
\midrule
CPP \texttt{charge} (deployed) & 0 & 0 & 0 & 0 & 0 & 0 & 0 & 0 & 0 & designated & no & 151,358 \\
\ML{} & 0 & 0 & 0 & 0 & 0 & 0 & 0 & 0 & 0 & designated & no & 98,084 \\
 
\bottomrule
\end{tabular}}
\end{table*}

\para{What the ablation shows} The two $C(M)$ rows differ from the baseline rows only in the nonce, and the \cls{order} column changes from 300 to 0. Order substitution is therefore not inherent to EIP-3009. It follows from the random nonce x402 clients draw, which the merchant cannot recompute. The \cls{redeemer} column changes only when the payee-bound variant is used. In the direct receive row the code-less merchant submits its own transaction. CPP and \ML{} add a recipient contract that checks the designated facilitator. They are the only tested configurations that keep facilitator settlement and close every class without a standing allowance, and both lose payee visibility (Proposition~\ref{prop:visibility}).

\para{What the baselines show} Three consequences matter for deployed protocols. First, the x402 exact scheme's authorization can be settled by any holder; Li et al.\ showed this preemption on a testnet~\cite{li2026fiveattacks}, and here it holds for the production token's code on every mandate. Second, Permit2's plain transfer lets the facilitator choose the recipient, so a compromised facilitator can redirect payments while remaining the only authorized redeemer. Third, approvals and permits bound only when the allowance may be created, not when it may be drawn, so a permit submitted inside the window can be drawn after it.

\section{The Boundary in Deployed Payments}
\label{sec:wild}
\subsection{Data}
We read the \texttt{AuthorizationUsed} events that Base USDC emits for every redeemed EIP-3009 authorization in 60 windows of 100 blocks each, spaced evenly by block height from block 45.5~million to 58.3~million (\nDateFrom{} to \nDateTo). Base's block interval fell during this period, so a window lasted between \nWinMinSec{} and \nWinMaxSec{} seconds. We sampled \nEvents{} events. For each, we decoded the signed fields from the calldata of the emitting transaction and accepted a decode only if the decoded payer and nonce equaled the event's indexed payer and nonce. This decoded \nNAuth{} authorizations (\nDecodeRate). We classified the remainder by receipt. \nNBatcher{} events came from one unverified batching contract whose runtime code calls \texttt{transferWithAuthorization} and whose calldata omits the nonce, which it therefore derives on chain. \nNCPP{} events came from transactions that emitted CPP logs. The rest (\nPathOther) passed through ERC-4337 bundles and other contracts.

To attribute payments to x402, we used the facilitator registry of the x402scan explorer at commit \nXCommit{}~\cite{x402scan}. Of decoded authorizations, \nXShare{} (\nXAttributed) were submitted by a registered x402 facilitator address, across \nXFacs{} facilitators; Coinbase's facilitator submitted \nXCoinbase{} of these. The amounts are micropayments: the median is \$\nValueMedian, and \nValueLtDollar{} are below one dollar.

\begin{figure*}[!t]
\centering
\includegraphics[width=\textwidth]{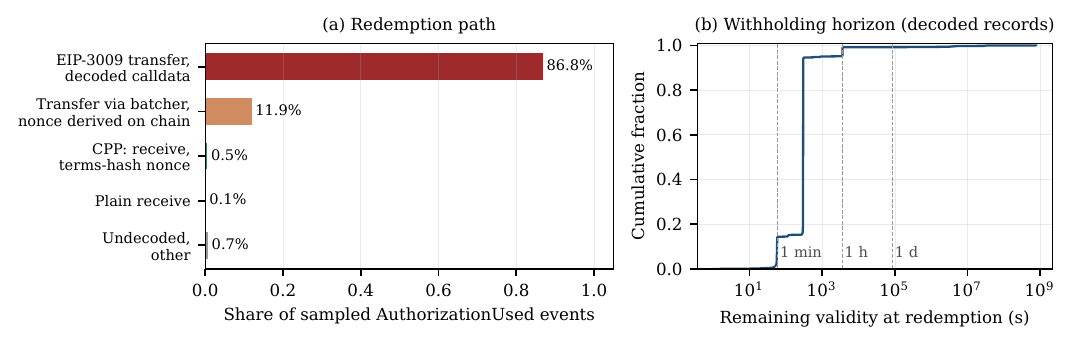}
\caption{Redeemed EIP-3009 authorizations on Base USDC, \nDateFrom{} to \nDateTo. (a) Redemption path of all \nEvents{} sampled events. (b) Remaining validity at redemption, for the \nNAuth{} decoded authorizations.}
\label{fig:base}
\end{figure*}

\subsection{Findings}
\para{Bearer redemption in the sample} Figure~\ref{fig:base}(a) shows that \nPathTransfer{} of events were decoded as \texttt{transferWithAuthorization}. Another \nPathBatcher{} went through the batching contract, which also uses the bearer variant. Together these two paths account for \nBearerShare{} of all \nEvents{} sampled events. Restricting each window to its first \nSliceSeconds{} seconds gives \nSliceBearer{} bearer events out of \nSliceAll{} sampled events (\nSliceBearerShare{}). The full-window and equal-duration shares are lower bounds within their samples because both denominators include unclassified events. Separately, among decoded events across the full windows, excluding submissions from Coinbase, the largest registered facilitator, leaves \nOtherBearer{} bearer events out of \nOtherDecoded{} (\nOtherBearerShare{}). Among x402-attributed authorizations, the payee-bound variant never appeared (\nXReceive). For the \nDirectShare{} of decoded authorizations submitted by a direct call to the token, where the sender is the token's caller, \nDirectThird{} were submitted by neither payer nor payee.

\para{Harm per class} With payee and amount bound, a party that redeems a bearer authorization cannot divert funds. It can preempt the designated facilitator, so the merchant sees a failed settlement for a paid request~\cite{li2026fiveattacks}. It can withhold the authorization and settle it later. And it can present one payment for a second, equal-priced request. We measure the time available for delay and the frequency of overlapping authorizations with matching terms.

\para{Withholding horizon} A holder of an unsettled authorization can redeem it until \texttt{validBefore}. The time from the observed redemption to \texttt{validBefore} is the further delay that holder could have imposed, a lower bound on its total discretion. Figure~\ref{fig:base}(b) shows a median of \nResMedian{} seconds, consistent with the five-minute default timeout of x402 clients. \nResGtHour{} exceeded one hour and \nResGtMonth{} exceeded 30 days. In \nValidAfterZero{} of authorizations \texttt{validAfter} was zero, so the deadline was the only temporal constraint. Long-lived authorizations are rare but present.

\para{Same-terms concurrency} Order attribution needs no on-chain commitment when the merchant records which nonce it received for which request. It fails when the merchant checks only payee and amount, a configuration Ling et al.\ found in most x402 deployments~\cite{ling2026freeriding}. To measure the exposure, we restricted each window to its first 20 seconds, so all windows have equal length (\nSliceRecords{} transfer-variant records). In \nConcurrent{} of these, another authorization in the same slice had the same payer, payee and amount and an overlapping validity interval. Such a pair is indistinguishable to a payee-and-amount verifier. A merchant needs either nonce bookkeeping or a commitment it can recompute, as in the $C(M)$ rows of Table~\ref{tab:matrix}.

\para{On-chain nonce derivation} We identified on-chain nonce derivation in \nCommitShare{} of sampled events, almost all from the batching contract. Calldata alone does not reveal how an off-chain signer chose a nonce. Its derived nonce supports attribution, but it is redeemed through the bearer variant, so anyone who obtains the signature can call the token directly. CPP transactions (\nPathCPP{} of the sample) were the only classified events in which we identified both a terms commitment and payee-bound redemption.

\section{Commitment-Bound Redemption}
\label{sec:design}
\subsection{Verifier Conditions}
Consider networks with distinct chain identifiers. Let $L$ redeem EIP-3009 authorizations of token $\tau_L$ on execution chain $\chi_L$. Write $M=(\chi_M,\bar M)$, where $\chi_M$ is the approved chain identifier and $\bar M$ holds the other mandate fields. Let $\N_L(M)=H(\mathsf{tag},\chi_M,L,\bar M)$, with Keccak-256 over an injective encoding of the fields and a constant domain tag. The trusted signer gets $\chi_M$ from the approved purchase and looks up $(\chi_L,\tau_L,L)$ in its trusted deployment configuration before producing $\texttt{ReceiveWithAuthorization}(p,L,v^\ast,t_a,t_b,\N_L(M))$ in $\tau_L$'s EIP-712 domain. $L$ satisfies the \emph{verifier conditions} if:
\begin{enumerate}
\item[V1] \emph{Commitment.} Each redemption forms $M'=(\chi_L,\bar M')$ from its execution chain and the caller's fields $\bar M'$, computes $n=\N_L(M')$ and calls \RWA{} on $\tau_L$ with arguments $(M'.p,L,M'.v^\ast,t_a,t_b,n,\sigma)$.
\item[V2] \emph{Checks.} Before calling the token, $L$ requires that $M'$ name $\tau_L$, that the caller be $M'.c^\ast$ (or $M'.c^\ast=\ast$), and that the block time lie in $[M'.t_0,M'.t_1]$.
\item[V3] \emph{Atomic forwarding.} In the same call, $L$ verifies that its balance grew by exactly $M'.v^\ast$, forwards exactly that amount to $M'.r^\ast$ and emits a log carrying $n$ and $M'.o^\ast$. Receipt and forwarding are the only movements of $\tau_L$ in this redemption.
\item[V4] \emph{No other path.} No other function of $L$ calls an authorization function of $\tau_L$ or moves tokens.
\end{enumerate}
CPP's tested zero-fee \texttt{charge} path uses a one-use signature for the specified amount and checks the operator and expiry. Its payment terms omit the start time $t_0$ of our general mandate. CPP deliberately has further paths (authorize, capture, void, refund) that hold funds in escrow, which is a different mandate semantics. \ML{} is a contract that satisfies V1--V4 and nothing else.

\subsection{Assumptions}
\begin{enumerate}
\item[A1] \emph{Token semantics.} A call of \RWA{} on $\tau_L$ with arguments $(p,\mathit{to},v,t_a,t_b,n,\sigma)$ succeeds \emph{only if} the caller is $\mathit{to}$, the time lies in $(t_a,t_b)$, $n$ is neither used nor cancelled for $p$, and $\sigma$ is $p$'s valid signature on the corresponding EIP-712 message in the domain for token $\tau_L$ and execution chain $\chi_L$. On success it marks $n$ used for $p$, moves exactly $v$ from $p$ to $\mathit{to}$ and calls no other contract for a code-less payer. The assumption covers the current USDC implementation; USDC is upgradeable.
\item[A2] \emph{Unforgeability.} The payer is a code-less EOA, and ECDSA is existentially unforgeable under chosen-message attack (EUF-CMA).
\item[A3] \emph{Collision resistance} of $H$ on injectively encoded inputs.
\item[A4] \emph{Signing discipline.} The payer's signer signs a \texttt{ReceiveWithAuthorization} naming $L$ only for a user-approved $M$ and after checking, immediately before issuing the signing request, that $\chi_M$, the configured deployment identifier $\chi_L$, the active wallet network and the signed \texttt{chainId} in the EIP-712 domain agree. It checks that the domain names $\tau_L$, derives the nonce and amount from the approved $M$, and requests a signature only when all checks pass.
\end{enumerate}

\subsection{Security Game and Theorem}
In the game $\mathsf{Exact}_{L}(\mathcal{A})$, the challenger holds the payer key. The adversary $\mathcal{A}$ adaptively requests mandates. The signer returns authorizations only for requests satisfying A4. Let $M_1,\dots,M_q$ be the issued mandates. The adversary controls every other account, deploys contracts and orders transactions. $\mathcal{A}$ wins if the final history contains a transaction that marks a nonce of $p$ used through a call with recipient $L$, and either (a) the nonce is not $\N_L(M_i)$ for any $i$, or (b) the effects attributable to that redemption violate the approved $M_i$ whose nonce it consumed.

\begin{theorem}[Mandate exactness]
\label{thm:soundness}
If $L$ satisfies V1--V4 and A1--A4 hold, then for every efficient adversary $\mathcal{A}$, $\Pr[\mathcal{A}\text{ wins }\mathsf{Exact}_L]\le\mathbf{Adv}^{\mathrm{euf\text{-}cma}}(\mathcal{B}_1)+\mathbf{Adv}^{\mathrm{cr}}_H(\mathcal{B}_2)$ for efficient $\mathcal{B}_1,\mathcal{B}_2$. Hence, except with negligible probability, $\Ex(M_i,\sigma_i)=\emptyset$ for every issued authorization.
\end{theorem}
\begin{proof}
Fix a transaction $T$ in the final history that marks a nonce $n$ of $p$ used through a call with recipient $L$. By A1 the caller of that token call is $L$. By V4 the call happens inside a redemption of $L$. By V1, $L$ supplies the execution chain field of $M'$ and the token call carries $(M'.p,L,M'.v^\ast,t_a,t_b,\N_L(M'))$. By A1 it succeeds only on a valid signature by $p$ on the typed message $\mu'$ for those fields, including the domain for $(\chi_L,\tau_L)$.

\emph{Case (a).} Suppose $\mu'$ was never signed. Then $\mathcal{B}_1$, which simulates the game with a signing oracle and outputs $(\mu',\sigma)$, forges, so this case occurs with probability at most $\mathbf{Adv}^{\mathrm{euf\text{-}cma}}(\mathcal{B}_1)$. Otherwise A4 gives $\N_L(M')=\N_L(M_i)$ and $M'.v^\ast=M_i.v^\ast$ for some approved $M_i$.

\emph{Case (b).} If $M'\neq M_i$, then $\mathcal{B}_2$ outputs the colliding preimages, with probability at most $\mathbf{Adv}^{\mathrm{cr}}_H(\mathcal{B}_2)$; the encoding includes the chain identifier and $L$. Otherwise $M'=M_i$, and the effect is allowed by $\Auth(M_i)$:
\begin{itemize}
\item V2 enforces the token, redeemer and both window bounds. A4 ties the approved chain to the signed domain, and A1 verifies that domain on $\chi_L$. The authorization is bound to the approved chain identifier, even when the payer uses the same key on other chains (\cls{chain}).
\item By V3 and A1 the net effect attributable to this redemption in $T$ is a transfer of exactly $v^\ast$ from $p$ to $r^\ast$ (\cls{payee}, \cls{overdraw}).
\item A1 marks $n$ used, so no later transaction consumes $\sigma_i$ (\cls{total}), and $\sigma_i$ creates no allowance (\cls{residual}).
\item A cancelled $n$ fails A1 (\cls{revoke}, if the cancellation is included first).
\item The emitted log carries $n=\N_L(M_i)$ and $o^\ast$. A verifier for $o\neq o^\ast$ recomputes $\N_L$ with order $o$, which differs from $n$ except with collision probability (\cls{order}).
\end{itemize}
No callback runs between the token's transfer to $L$ and the forward (A1, V3), so no reentrant call observes or spends the intermediate balance.
\end{proof}

\para{Outside the theorem} The theorem guarantees that the chain enforces the mandate the signer committed to. It does not guarantee that the mandate reflects the user's intent; that is the job of the trusted signer and of defenses against pre-signature manipulation~\cite{louck2026whisper}. Forks that reuse a chain identifier share this network binding. A1 and A2 restrict payers to code-less accounts. For ERC-1271 smart accounts and EIP-7702-delegated accounts, soundness can depend on the account's validation code~\cite{erc1271,eip7702}. Revocation races with settlement, as for every row of Table~\ref{tab:binding}.

\subsection{\ML{}}
\ML{} is a contract with no owner, no upgrade path and no storage, deployed once per token. Its single \texttt{settle} function implements V1--V3 (Fig.~\ref{fig:overview}). It reverts on a wrong token, a wrong caller or a time outside the window. It fills the chain field from \texttt{block.chainid}, derives the nonce, calls \texttt{receiveWithAuthorization}, checks the exact balance increase, forwards to the merchant and emits $\mathsf{Settled}(n,o^\ast,p,r^\ast,v^\ast)$. The signer checks the chain before signing under A4. A merchant accepts payment for order $o$ when it observes $\mathsf{Settled}$ from $L$ whose nonce equals $\N_L(M)$ recomputed from its own quote. This rejects a buyer's self-signed mandate for the same order at a lower price, because the recomputed nonce differs. For x402, \ML{} is a new scheme in which \texttt{PaymentRequirements} name the lock, the merchant, the facilitator and a resource digest. In AP2, the cart mandate hash becomes $h$, and the credentials provider computes $\N_L(M)$ and produces the signature.

Compared with CPP, \ML{} covers only the immediate-payment case. It holds no escrow state, needs no operator registration, fee configuration or token store, and commits the window start and the cart digest in addition to the order. These restrictions make it shorter to audit and cheaper to run (Section~\ref{sec:cost}). CPP should be preferred when a payment needs capture or refunds.

\section{Evaluating \ML{}}
\label{sec:locketal}
\para{Mutation sweep} We ran \nSweepTotal{} settlement attempts on the deployed USDC code, each on a freshly signed mandate, in three modes:
\begin{enumerate}
\item[(i)] the signed window equals the mandate window;
\item[(ii)] the signed window is unbounded, so the lock's own window check decides;
\item[(iii)] no facilitator is designated.
\end{enumerate}
Each attempt mutated none or one of the nine mandate fields, including the token, and used the facilitator or a stranger as caller at a time within two days. \ML{} rejected all \nSweepMutated{} attempts with a mutated mandate (\nSweepMutAccepted{} accepted). Of the \nSweepUnmodified{} attempts with an unmodified mandate, it accepted the \nSweepAccepted{} whose caller and time satisfied the mandate. Payer debits were exact and the lock's balance stayed zero (\nSweepMismatch{} disagreements with the mandate predicate).
\para{Fuzzing and spoofing} A property-based Foundry test asserts the same predicate, plus failure of a second settlement, over fuzzer-chosen mandates, mutations, callers, times, window modes and facilitator settings, and passed 10,000 runs. Two targeted tests check spoofing. A caller-supplied fake token is rejected. A buyer's self-signed one-unit mandate for a victim's order settles, as intended, but the merchant verifier does not accept it as payment for the quoted order.
\para{Standard signers} For 50 mandates we exported the credential as an \texttt{eth\_signTypedData\_v4} document, the format wallets accept for EIP-712 requests. We signed each with two implementations, Foundry's \texttt{cast} and the Python \texttt{eth-account} library 0.13.7~\cite{ethaccount}. All signatures settled (\nWalletCast{} and \nWalletEA).
\para{Approved network} We independently varied the buyer's approved network, the configured deployment, the wallet's selected network and the network in the signing domain. With two possible identifiers for each input, a mock wallet provider received a signing request in the two cases where all four agreed. The other fourteen combinations and a domain that omitted the network from its signed fields yielded no signing request. In a separate contract test, the commitment matched an independently encoded value for the approved network and changed when only the execution chain identifier changed.

\begin{figure*}[t]
\centering
\includegraphics[width=0.91\textwidth]{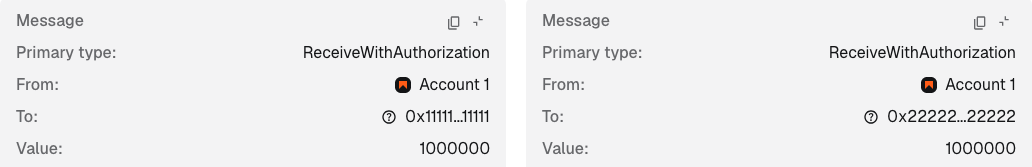}
\caption{Message details from the MetaMask approval screen for the same order. Left, the direct request shows the merchant (\texttt{0x11111...11111}) in the \texttt{To} field. Right, the mediated request shows the lock (\texttt{0x22222...22222}). Both display the same amount in token base units. The supplement includes the full screens, nonce views and unsigned requests.}
\label{fig:wallet-screen}
\end{figure*}

\para{Wallet approval screen} In two fresh browser profiles, we tested the same three request conditions using expired synthetic authorizations in the released MetaMask~13.33.0 extension~\cite{metamaskpreview}. For the first pair, we held the account, amount, order commitment and signing domain fixed and changed only the recipient. MetaMask displayed the merchant address for direct payment and the lock address for mediated payment (Fig.~\ref{fig:wallet-screen}). For the third request, we changed only the order in the mandate and recomputed its commitment. The nonce changed on the wallet screen. The recipient and amount stayed the same, and neither mediated screen named the merchant or showed the order label. We canceled all six requests. This measures what one wallet version displayed on a local network with external security lookups unavailable.

\subsection{Cost}
\label{sec:cost}
We measured gas with each honest settlement executed as its own transaction (Foundry isolation mode), over 50 mandates per configuration, with payer and merchant already holding balances. Gas includes the 21,000 intrinsic cost and calldata and is reported before refunds. We priced the L1 data fee by calling Base's fee oracle at block 58,495,000 on a signed transaction carrying the exact calldata. The L2 base fee was \nBaseFeeGwei{} gwei, and we used the 17 November 2026 ETH price of \$\nEthUsd{}~\cite{coingecko}.

\ML{} uses \nGasLock{} gas against \nGasReceive{} for an unbound payee-bound transfer (+\nGasOverheadPct), \nGasHardened{} for the hardened Permit2 proxy and \nGasCPP{} for CPP's \texttt{charge}. The overhead comes from the second token transfer and the longer calldata. At the pinned block a settlement costs \$\nCostLock{} with \ML{}, against \$\nCostTransfer{} for the bare x402 transfer and \$\nCostCPP{} for CPP. That is \$\nCostDelta{} more than a payee-bound transfer. The facilitator pays it. Relative to the measured median payment of \$\nValueMedian{}, settlement costs \nCostTransferPctMedian{} of the payment without binding and \nCostLockPctMedian{} with \ML{}. For sub-cent payments the binding is not free, and batching settlements is the natural next optimization.

\section{Discussion}
\label{sec:discussion}
\para{Implications for protocol design} Agent commerce protocols should treat the wallet signature, not the mandate credential, as the enforcement boundary, and choose a configuration whose redemption rules enforce the full mandate, including the exact amount and the permitted number of payments. For x402, two changes close the measured gaps without new token features. Clients can commit the requirements in the nonce, so merchants can recompute it. Facilitators can redeem through a commitment-checking recipient contract. For AP2's crypto path, the cart mandate's hash should enter the signed payment authorization, not only the credential chain.
\para{Implications for wallets} When a separate recipient contract forwards the payment to a code-less merchant, the EIP-3009 request names the forwarding contract (Proposition~\ref{prop:visibility}). Wallets can close this by recomputing $\N_L(M)$ from a mandate presented with the request and displaying the merchant, amount, order and facilitator it commits to.
\para{Smart accounts} ERC-7710 delegations with caveat enforcers can express the same constraints for smart accounts~\cite{erc7710,erc7715}, and ERC-4337 and EIP-7702 enable related scoping~\cite{erc4337,eip7702}. These need a smart account and a correct composition of enforcers. Our analysis covers EOA payers signing token-level authorizations.
\para{Operational risks} A designated facilitator can refuse to settle, which delays the merchant but cannot move funds; a short window bounds the delay. A lock is a single contract that the token's administrator could blacklist, which would halt settlements through it. Merchants should treat settlements as final only after the chain's finality rule. Identical repeat purchases need distinct order identifiers. The $\mathsf{Settled}$ event discloses the order identifier, which deployments can replace by a salted commitment.
\para{Future work} An end-to-end study could test whether shopping agents obtain the right approval and complete the corresponding purchase. On a local test network, the same purchase tasks could be paired across standard wallet prompts and mandate-aware prompts while holding the agent, quotes and settlement checks fixed. Test cases would include matching quotes, revised prices, changed orders and expired approvals. Outcomes would include exact completion, requests to proceed on mismatched terms, and requests for clarification. This would measure how approval information affects purchase outcomes before and during settlement.

\section{Threats to Validity}
\label{sec:validity}
\para{Internal} The same authors wrote the derivation, the attacks and the oracle. To limit circularity, the oracle reads only balances and logs, the honest check is two-sided, and the \cls{order} outcome is scored by one verifier applied to every row. A \checkmark{} cell is a universal claim, which one scripted attack per cell cannot prove. The derivation in Appendix~\ref{app:derivation} and, for \ML{}, the sweep, fuzzing and Theorem~\ref{thm:soundness} carry those claims.
\para{Construct} Bearer redemption is exposure, not harm: we do not observe preemption or misattribution. Same-terms concurrency measures what a payee-and-amount verifier cannot distinguish; merchants that track nonces are not affected. The withholding horizon is a lower bound on a holder's discretion, since signing time is not on chain.
\para{External} The reported shares describe redeemed USDC authorizations in the \nWinSampled{} sampled Base windows. Each event has equal weight, so busier windows contribute more observations. x402 attribution relies on a public registry and therefore misses unregistered facilitators. \nPathOther{} of events remain unclassified. The batching contract is unverified; we classify it from its bytecode and calldata. The two Permit2 proxies are our implementations; the deployed x402 Permit2 proxy may differ. The EVM experiments use fresh contract state and a supplied approval. They measure settlement behavior. How often a shopping agent obtains the intended approval across a complete purchase remains open.

\section{Related Work}
\label{sec:related}
\para{Agent payment protocol security} Li et al.\ formalize x402, demonstrate settlement preemption, replay across the HTTP--chain boundary and header manipulation, and propose facilitator-bound settlement~\cite{li2026fiveattacks}. Ling et al.\ identify cross-resource substitution and settlement races and propose extending the signed schema with a request hash~\cite{ling2026freeriding}. Wang et al.\ build x402scope, find facilitator flaws and measure facilitator concentration in 2025 traffic~\cite{wang2026http402}. Jiang et al.\ verify x402, MPP, AP2 and ACP in the Tamarin prover~\cite{jiang2026formal}. Zhang et al.\ systematize agent-to-agent payments~\cite{zhang2026sok}. Table~\ref{tab:related} positions our work. We contribute a model of the wallet-side boundary that covers all mechanisms in one set of classes, validate it on deployed bytecode with an ablation, measure configuration choices rather than flaws, and prove the binding pattern.

\begin{table}[t]
\caption{Positioning against the closest work.}
\label{tab:related}
\centering\footnotesize
\setlength{\tabcolsep}{2.5pt}
\resizebox{\columnwidth}{!}{%
\begin{tabular}{@{}lccccc@{}}
\toprule
Work & Wallet primitives & Deployed code & Mainnet data & Binding & Proof \\
\midrule
Li et al.~\cite{li2026fiveattacks} & 2 & testnet & -- & caller & protocol \\
Ling et al.~\cite{ling2026freeriding} & 1 & SDKs & census & schema & bounds \\
Wang et al.~\cite{wang2026http402} & 1 & facilitators & 2025 & -- & -- \\
Jiang et al.~\cite{jiang2026formal} & -- & -- & -- & -- & Tamarin \\
CPP~\cite{cpp} & 4 & deployed & -- & terms+caller & -- \\
This work & 12 configs & Base bytecode & 2026 & terms+caller & game \\
\bottomrule
\end{tabular}}
\end{table}

\para{Mandate integrity in AP2} Debi et al.\ red-team AP2 through prompt injection~\cite{debi2026whispers}. Aviv et al.\ show that mandate signatures can omit the operational context from which the agent built the mandate~\cite{aviv2026beyondmandate}. Louck et al.\ propose A-VIP, which binds a credentials-provider token to the cart, the redeeming merchant and a nonce~\cite{louck2026whisper}. These works protect what the user signs. Ours protects what the signature permits once it leaves the wallet. A-VIP's cart binding provides the digest $h$ that a commitment-bound authorization carries.

\para{Settlement designs} A402 binds payment to service execution with adaptor signatures and trusted hardware~\cite{a402}. The Agentic Settlement Protocol specifies escrowed, refundable settlement with offer-bound authorizations~\cite{asp2026}. CPP~\cite{cpp} deploys authorize-and-capture escrow. Our verifier conditions explain the binding checks on CPP's charge path, and \ML{} isolates the minimal mechanism.

\para{Agent tool-use security} AgentDojo~\cite{agentdojo}, CaMeL~\cite{camel} and Progent~\cite{progent} constrain what an agent decides or which tools it may call. They do not address what a signed payment credential permits after it leaves the agent.

\section{Conclusion}
When an AI agent pays in stablecoins, the commerce protocol states the mandate, but the token enforces only the wallet's signed fields. We defined the resulting authority gap and compared nine classes of violations across twelve configurations. The derivation predicted every executed trial on deployed bytecode. In the sampled Base windows, we identified bearer redemption for \nBearerShare{} of redeemed USDC authorizations. Under the assumptions of Theorem~\ref{thm:soundness}, a recipient contract that checks the committed terms, caller and time and forwards the exact amount once makes the signature mandate-exact. When a separate contract forwards to a code-less merchant, its address occupies the payee field shown by the wallet. The commerce protocol and the wallet can therefore share one enforced boundary with signature formats that wallets already support, provided the signer that computes the commitment is trusted to show the mandate.

\section*{Ethical Considerations}
The measurement reads public blockchain data and reports aggregates only; it does not link addresses to persons. All attack trials ran against local copies of contract code, and no transaction was sent to a live network. The gaps we report follow from public specifications and from issues already disclosed for x402~\cite{li2026fiveattacks,ling2026freeriding}; we found no new vulnerability in a specific deployment.

\section*{Artifact Availability}
The reviewer artifact repository is \url{https://anonymous.4open.science/r/PaperA-Code-Policy-Gate-1752/}. The supplement provides the decoded measurement records, receipt classification counts, pinned facilitator registry, and scripts to recompute the measurement and summarize recorded experiment outputs. It includes \ML{}, pinned token bytecode with digests, local tests of authorized settlement and the chain commitment, and the wallet preview fixture and screen captures. The validation matrix, mutation sweep, gas benchmarks, and signer comparisons are supplied as recorded results.

\appendices
\section{Derivation of Table~\ref{tab:binding}}
\label{app:derivation}
Each row lists how the signed fields are filled from $M$, the checks on the redemption path and a witness history for every open class.
\para{Approve (exact allowance / unlimited)} The payer approves $c^\ast$ for $v^\ast$ (respectively unlimited). \texttt{transferFrom} checks only the allowance. Witnesses: $c^\ast$ transfers to itself (\cls{payee}), after $t_1$ (\cls{late}) or for an equal-priced second order (\cls{order}). The unlimited row additionally admits $v^\ast+1$ (\cls{overdraw}), a second draw (\cls{total}) and a draw 30 days later (\cls{residual}).
\para{Permit} The signature sets an allowance of $v^\ast$ for $c^\ast$ with deadline $t_1$. The deadline bounds only allowance creation. Witnesses: as for exact approval, with the permit submitted inside the window and the draw made after $t_1$. Revocation burns the permit nonce with a self-submitted zero-value permit.
\para{EIP-3009 transfer / receive} $(p,r^\ast,v^\ast,t_0-1,t_1+1,n)$ with a random $n$. Transfer witnesses: any account submits (\cls{redeemer}); an equal-priced second order is accepted by a payee-and-amount verifier (\cls{order}). Receive requires caller $=r^\ast$, so \cls{redeemer} is bound with $c^\ast=r^\ast$; \cls{order} is open as for transfer. Revocation is \texttt{cancelAuthorization}.
\para{Permit2 transfer} Signature over $(\tau,v^\ast,c^\ast,n,t_1)$ after an unlimited Permit2 approval. The spender chooses the recipient (\cls{payee}); there is no order field (\cls{order}). Revocation invalidates the unordered nonce.
\para{Permit2 witness and hardened proxies} The witness fixes $r^\ast$ and $t_0$; the x402-style proxy accepts any caller (\cls{redeemer}) and has no order field (\cls{order}). The hardened witness adds $c^\ast$, with a caller check, and $o^\ast$; it binds all classes and keeps a standing Permit2 allowance.
\para{$C(M)$ rows} As the EIP-3009 rows with $n=H(\chi,M)$. The verifier recomputes $n$ for the claimed order, which closes \cls{order}; transfer still admits any caller.
\para{CPP charge and \ML{}} Recipient contract with a terms-hash nonce and a caller check. CPP's \texttt{PaymentInfo} carries operator, receiver, amount, expiries and salt; \ML{}'s mandate carries all fields of $M$. Both bind all classes (Theorem~\ref{thm:soundness} for \ML{}). Revocation is \texttt{cancelAuthorization} on the derived nonce.

\bibliographystyle{IEEEtran}
\bibliography{references}
\end{document}